\documentclass[a4paper,preprint]{article}
\usepackage{graphicx}
\begin{document}

\date{\today}

\title{Giant fluctuations and gate control of the $g$-factor in  InAs Nanowire Quantum Dots}

\maketitle

\author{S.~Csonka\footnote{Email:\ \texttt{szabolcs.csonka@unibas.ch}\ \ \
  Web:\ \texttt{http://www.unibas.ch/phys-meso}}, L.~Hofstetter, F.~Freitag, S.~Oberholzer and
  C.~Sch\"{o}nenberger
  \\\textit{Department of Physics, University of Basel, Klingelbergstr. 82, CH-4056 Basel, Switzerland}\\}
\author{T.~S.~Jespersen, M. Aagesen and J.~Nyg{\aa}rd\\
\textit{Nano-Science Center, Niels Bohr Institute, University of Copenhagen, Universitetsparken 5, DK-2100 Copenhagen,
Denmark}}

\begin{abstract}

We study the  $g$-factor of discrete electron states in InAs nanowire based quantum dots. The $g$ values are determined
from the magnetic field splitting of the zero bias anomaly due to the spin $1/2-$Kondo effect. Unlike to previous
studies based on 2DEG quantum dots, the $g$-factors of neighboring electron states show a surprisingly large
fluctuation: $g$ can scatter between 2 and 18. Furthermore electric gate tunability of the $g$-factor is
demonstrated.\end{abstract}

Due to the precise structural control, semiconductor nanowires (NW) provide a new class of nanoscale building blocks
for a broad range of disciplines like quantum optics, electronics, nanosensing and biotechnology~\cite{Lieber2007}.
Recently semiconductor NWs also attract increasing attention in the field of quantum transport: single electron
transistor behavior~\cite{Franceschi2003}, Pauli spin blockade~\cite{Pfund2007} and the Kondo
effect~\cite{Jespersen2006} have already been observed in NW based devices. Furthermore, new physical phenomena were
also demonstrated, e.g. $\pi$-junction~\cite{vanDam2006} or the interplay of Andreev reflection and Kondo
effect~\cite{Jespersen2007,Eichler2007}, which are not accessible in conventional GaAs heterotructures. In this Letter
we report on a surprising behavior of InAs NW based quantum dots (QD). We investigate the $g$-factor of the dot
electron states, which is the key parameter for the manipulation of spin information at a single electron level. The
$g$-factor shows an order of magnitude variation for neighboring discrete electron states, which highly exceeds
previous observations in semiconductor based quantum dots~\cite{Folk2001,Hanson2007}. Moreover, strong gate tunability
of the $g$ value of individual charge states is also demonstrated.

\begin{figure}[t!]
\centering
\includegraphics[width=100mm]{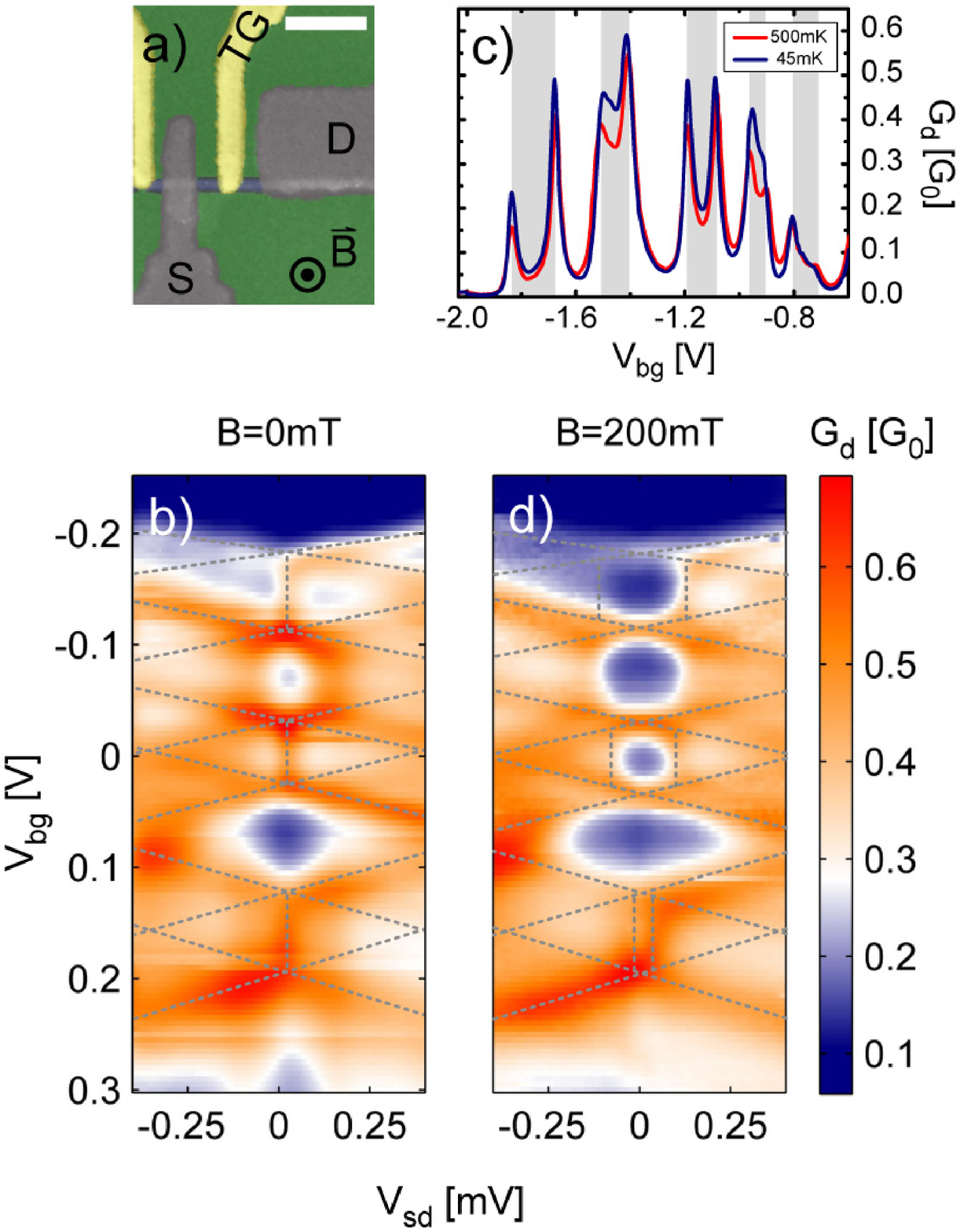}
\caption{\it (a) SEM picture of the device: The quantum dot forms between the two Ti/Al ohmic contacts (S,D) with Ti/Au
topgate (TG) deposited on top of it. The nanowire (blue) has a diameter of $80\,$nm. The bar is $0.5\,\mu$m . (b,d)
Colorscale plots of the differential conductance vs. backgate and source drain voltage measured at $V_{tg}=0.07\,$V.
The stability diagram at $B=0\,$T (b) clearly shows the even/odd filling of the quantum dot. There is a Kondo ridge at
zero bias voltage in every second Coulomb diamond (dotted lines at $V_{sd}=0$). At $B=200\,$mT (d) the Kondo ridges
split up differently. (c) Differential conductance versus backgate voltage at $T=45\,$mK and $500\,$mK. The conductance
increases inside every second Coulomb diamond for decreasing temperature  (gray regions) in agreement with Kondo
physics. Measured at $V_{tg}=0.06\,$V and $V_{sd}=0\,$mV. } \label{Fig1.fig}
\end{figure}

The high quality InAs nanowires were grown by molecular-beam epitaxy~\cite{Jespersen2006}. The nanowires were then
brought into IPA solution by short sonication of the growth substrate. The wires were deposited by spinning a droplet
of the solution on a highly doped Si substrate with $400\,$nm insulating SiO$_2$ cap layer. After locating the
nanowires by SEM, the devices were fabricated by standard e-beam lithography technique.  We define ohmic source and
drain contacts for transport measurements (S,D) and a top gate electrode (TG) for local gating (see
Fig.~\ref{Fig1.fig}a). The Ti/Au ($10/100\,$nm) TG electrode is isolated from the nanowire by the native surface oxide
layer~\cite{Shorubalko2006}. Before the evaporation of the Ti/Al ($10/100\,$nm) ohmic contacts this oxide layer is
removed by gentle Argon sputtering. The transport measurements were performed with standard lock-in technique in a
dilution refrigerator at a base temperature of $40\,$mK in perpendicular magnetic field. The electron density of InAs
wires can be strongly varied by applying voltage on the back gate electrode. Based on the backgate dependence the
nanowires have n-type charge carriers. When the back gate potential is decreased, barriers are generated at the
contacting source and drain electrodes, and a QD forms in the middle of the wire
segment~\cite{Franceschi2003,Jespersen2006}. In our device architecture the topgate provides an additional knob to
modify the shape of the confinement potential of the QD besides the back gate.

A typical stability diagram measured on such a QD is plotted in Figure~\ref{Fig1.fig}b. Since the dot is in a rather
open state the differential conductance exceeds the value of $\sim0.7\,$G$_0$ (G$_0=2e^{2}/h$) and the borders of the
Coulomb blockade diamonds are rather blurred. From the size of the diamonds (see the dotted lines) the estimated
charging energy is in the range of $0.5-1\,$meV, which is consistent with the length of the nanowire segment of
$400\,$nm~\cite{Jespersen2006}. The subsequent Coulomb diamonds have alternating sizes (smaller / larger /smaller ...),
and in the smaller diamonds the differential conductance shows a pronounced increase at zero source drain bias voltage
(see dotted lines at $V_{sd}=0$ in Fig.~\ref{Fig1.fig}b). This type of even-odd dependence of the stability diagram is
a strong indication of the spin$-1/2$~Kondo effect~\cite{Goldhaber-Gordon1998}, which was analyzed in InAs nanowire QDs
by Jespersen~$et$ $al.$ in detail~\cite{Jespersen2006}.

\begin{figure}[t!]
\centering
\includegraphics[width=80mm]{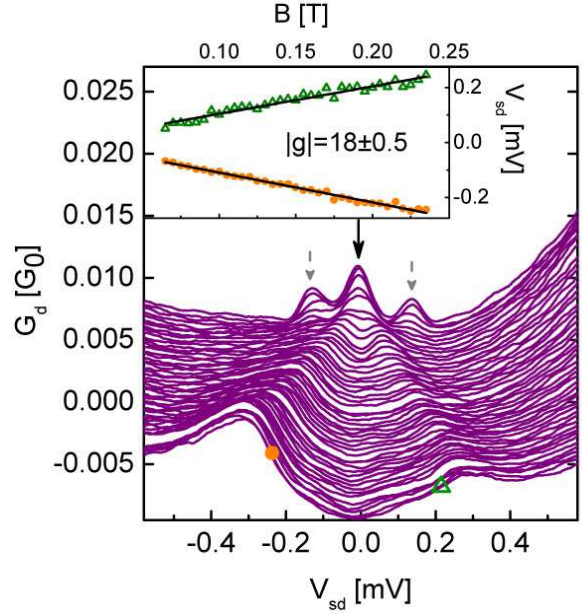}
\caption{\it Magnetic field splitting of the Kondo zero bias anomaly. (main panel): Differential conductance vs. source
drain voltage at different perpendicular magnetic field values $B=0,  5, 10, ..., 240\,$mT (back to front). Measured at
$V_{bg}=-2.64\,$V and $V_{tg}=0.08\,$V. The curves are shifted for clarity. Note that the two side peaks (gray arrows)
are superconducting features induced by the Ti/Al electrodes. (inset): The position of the inflection points of the
$G_d(V_{sd})$ curves from the main panel (orange dot and green triangle) as a function of the magnetic field. Linear
fit (line) with the extracted $|g|$-factor. } \label{Fig2.fig}
\end{figure}

If the dot contains an odd number of electrons the highest occupied electron state is filled only by a single electron.
Due to the strong coupling to the leads, the spin of this unpaired electron is screened by electrons of the leads
forming a many-body state, the so-called Kondo cloud, which gives rise to the enhancement of conductance at zero bias
voltage. The temperature dependence of the slice of the stability diagram at $V_{sd}=0$ is also consistent with the
Kondo effect (see Fig.~\ref{Fig1.fig}c). Inside the diamond with even number of electrons the differential conductance
decreases for decreasing temperature, while an increase is observed in odd diamonds (gray regions), where Kondo
correlations take place. The spin$-1/2$ Kondo physics has a characteristic magnetic field dependence. The Kondo ridge
at zero bias voltage splits up in a magnetic field similar to the Zeeman-splitting according to $\Delta V =
2g\mu_{B}B/e$, where $\mu_{B}$ is the Bohr magnetron, $e$ is the electron charge and $g$ is the effective $g$-factor of
the highest discrete electron state occupied with an unpaired electron~\cite{Meir1993, Cronenwett1998}. The magnetic
field dependence of a zero bias ridge is shown in Figure~\ref{Fig2.fig}. The zero bias anomaly peak (black arrow)
splits up linearly with magnetic field in agrement with Kondo physics (see inset). Note that two side peaks at low $B$
field (gray arrows) are related to the superconducting electrodes and they are completely suppressed by a field of
$B>50\,$mT. They will not be considered further in this paper~\cite{Jespersen2007}. Since the magnetic field splitting
of the zero bias Kondo ridge is proportional to the Land\'{e} $g$-factor of electron state, one can evaluate the
absolute value of the $g$-factor of the highest occupied discrete electron state based on the magnetoconductance data
(see Fig.~\ref{Fig2.fig}). There are different methods in the literature for the precise determination of the $|g|$
value from the Kondo splitting. In the first work the positions of the maximums of the splitted Kondo peak were
used~\cite{Meir1993,Cronenwett1998}, which yields a slight overestimation (few percent) according to later
studies~\cite{Kogan2004}.  In our work we will determinate $|g|$ from the magnetic field dependence of the two
inflection points (see orange dot and green triangle for the highest $B$ field in the main panel of
Fig.~\ref{Fig2.fig})~\cite{Paaske2004}, which provides $\sim5\%$ smaller values. The position of the inflection points
are plotted as a function of the B field in the inset of Fig.~\ref{Fig2.fig} and their difference is fitted by $\Delta
V = 2|g|\mu_{B}B/e$.

\begin{figure}[t!]
\centering
\includegraphics[width=80mm]{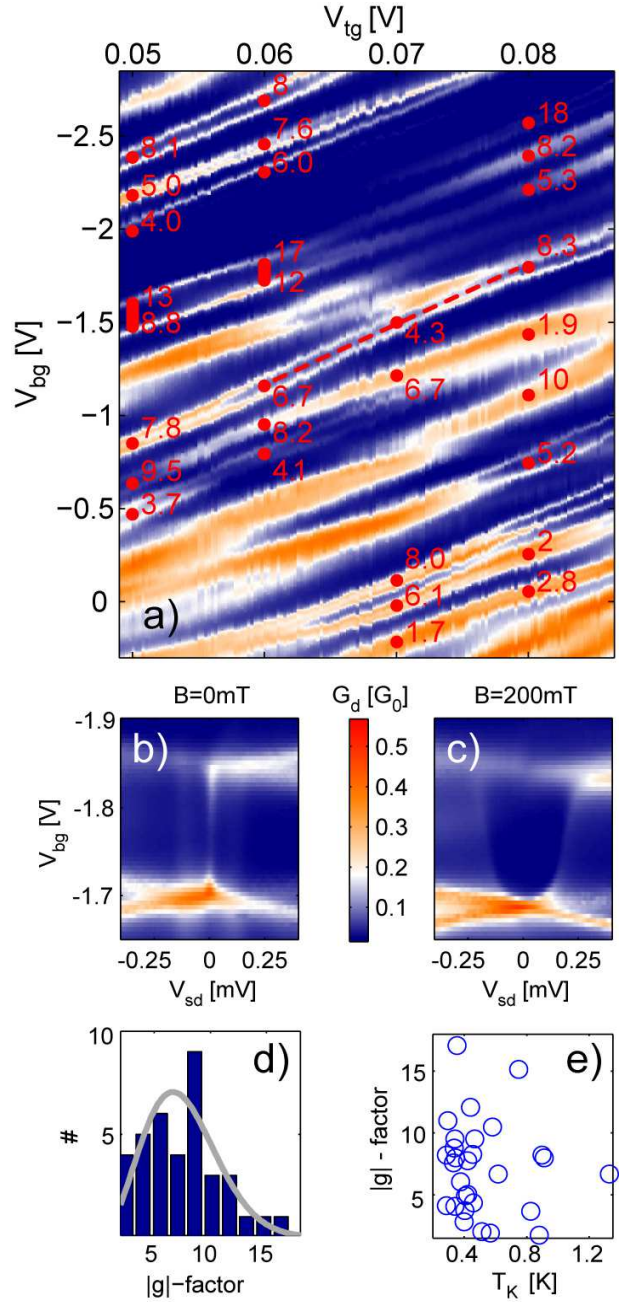}
\caption{\it (a) Summary of the evaluated $|g|$ values. Colorscale plot of the differential conductance vs. backgate
voltage and TG voltages (at $V_{sd}=0\,$V and $B=0\,$T) with the absolute value of the $g$-factors determined from the
magnetic field splitting of the Kondo ridges. (b,c) Stability diagram of a Coulomb diamond with a Kondo zero bias
anomaly, where the magnetic field induced splitting is changing inside the diamond. Measured at $V_{tg}=0.6\,$V. (d)
Distribution of the evaluated $|g|$-factors (bar) and a parameterless fit (line), see text. (e) Extracted $|g|$ values
vs. estimated Kondo temperature from the full width at half maximum of the Kondo zero bias anomaly at $B=0\,$T
(FWHM=$2k_{B}T_K/e$). } \label{Fig3.fig}
\end{figure}

In the following we analyze how the effective $g$-factor of individual discrete electron states behaves when the
confinement potential of the QD is changed with the gate electrodes. Already the stability diagram measured at
$B=200m$T in Fig.~\ref{Fig1.fig}d shows that the magnetic field splitting of the Kondo ridge in subsequent odd Coulomb
diamonds varies strongly. The extracted $|g|$ values of these three electron states are $\sim8$, $6.1$ and $1.7$. Thus
adding only two extra electrons onto the dot can change the $g$-factor by over $100\%$. Fig.~\ref{Fig3.fig}a summarizes
the evaluated $|g|$ values. The colorscale plot shows the differential conductance vs. topgate and backgate voltages.
Higher conductance regions appear along diagonal lines since the backgate voltage induced shift of the electron levels
of the dot is compensated by opposite tuning of the topgate. At four different TG voltages the coordinates of the
diamonds where Kondo ridges were observed (similar ones as in Fig.~\ref{Fig1.fig}b) are plotted by circles.  The
neighboring numbers are the absolute value of the $g$-factors evaluated similar to the way in Fig.~\ref{Fig2.fig}.
$|g|$ does not show a systematic behavior as a function of the backgate voltage, but rather changes in a random
fashion. The extracted $|g|$ values vary in a broad range (see also the distribution in Fig.~\ref{Fig3.fig}d) from the
lowest value of $1.7$ even up to $18$, which exceeds the InAs bulk value of $|g|=14.7$. By modifying the backgate and
topgate voltage simultaneously along the Coulomb blockaded regions (blue stripes)  the confinement potential of the dot
changes while its charge state is preserved. Such a tuning of the gate voltages can also lead to marked $g$-factor
variations.   For instance along the dashed line in Fig.~\ref{Fig3.fig}a the $|g|$ varies by almost a factor of two,
while the number of electrons is kept fixed on the dot. Such a gate tunability of the $g$-factor of single electron
state provides an efficient way for selectively addressing electron spins. Furthermore, in some of the Coulomb diamonds
the magnetic field splitting of the Kondo ridge is clearly $not$ constant over backgate voltage. As it is seen in
Fig.~\ref{Fig3.fig}b-c the splitting of the Kondo ridge varies more than $30\%$ between the two sides of the diamond at
$B=200$mT. Thus changes of the $g$-factor are visible even inside a single Coulomb diamond. Summarizing, the $g$-factor
fluctuates by an order of magnitude for neighboring electron states and it can be strongly gate-tunable even in the
same charge state. We investigated the relation between the observed $g$ values and Kondo temperature ($T_K$) estimated
from the full width at half maximum of the Kondo ridges~\cite{Cronenwett1998}. As it is shown in Fig.~\ref{Fig3.fig}e,
no correlation was found between $g$ and $T_K$. Therefore we conclude that the level-to-level fluctuation of the
$g$-factor is not related to Kondo physics.

The rather large $g$-factor of bulk InAs  ($g = -14.7$) is due to the presence of strong spin-orbit coupling and the
contribution of the orbital degree of freedom. If the electrons are confined in a small constriction the large $|g|$
value decreases even down to the free electron value of $2$ because of the quenching of the angular
momentum~\cite{Kiselev1998,Pryor2006}. This phenomenon was reported by Bj\"{o}rk~\textit{et al.} for InAs
nanowires~\cite{Bjork2005}. They studied quantum dots formed between two (few nanometer thin) InP tunnel barriers.
Decreasing the distance between the barriers to $8$nm the $|g|$ value decreased down to $\simeq 2$. However, they did
not observe measurable changes in $|g|$ varying the number of electrons on the quantum dot. The size of the QD in our
study is relative large, $L\simeq 350$nm and the occupation number is also significantly higher. Therefore, a
considerable orbital angular momentum is preserved. This is supported by the average value of the measured $g$-factors
of $\langle|g|\rangle = 7.4\pm3.6$. The elastic mean free path and spin orbit (SO) length ($l_e\approx80$nm,
$l_{SO}\approx125$nm~\cite{Fasth2007}) are significantly smaller than the size of the dot, thus the electrons cover
diffusive-like trajectories and several SO scattering events are expected to take place in the dot.

Although the strong fluctuation of the $g$ value of neighboring discrete electron states is surprising for
semiconductor nanostructures, it was observed in metallic nanoparticle based QDs in the presence of spin orbit
interaction~\cite{Petta2001}. The $g$-factor of the discrete electron levels for e.g. silver nanoparticles (diameter of
$\simeq5-10$nm) varies between $0.25-1$. The $g$-factor of a particular state contains contributions from the SO matrix
elements with all other discrete levels. Since the precise nature of the wave function of electron states is
fluctuating and the level spacing is also varying the SO induced correction exhibits fluctuation. The observed
fluctuation agrees quantitatively with random matrix theory (RMT) calculations~\cite{Matveev2000,Brouwer2000}, which
describe the SO interaction with the parameter $\lambda=(\hbar\pi /\tau_{SO}\delta)^{0.5}$ ($\tau_{SO}$ and $\delta$
are the SO scattering time and the level spacing, respectively).  In our InAs NW measurements a rough estimation of the
SO strength is $\lambda\approx0.4$ based on $\delta\approx0.1\,$meV and $\tau_{SO}\approx100\,$ps (bulk
value)~\cite{Murdin2005}. This is close to the range of $\lambda\approx0.7-10$, where $g$ fluctuation was observed in
metallic grains~\cite{Petta2001}. RMT provides an analytic expression for the distribution of the $g$-factor in the
strong SO coupling regime ($\lambda\gg1$):
\begin{equation}
P(g)= 3 \left(\frac{6}{\pi}\right)^{1/2} \frac{g^2}{\langle g^2\rangle^{3/2}} \exp {
  \left ( -\frac{3g^2}{2  \langle g^2\rangle}\right)},
 \label{Eq1}
\end{equation}
where  $\langle g^2\rangle$ is the mean square value of the distribution. For our experimental distribution function of
$|g|$ (see bar graph in Fig.~\ref{Fig3.fig}d) $\langle g^2\rangle=67$ . Using this mean square value, the RMT predicted
distribution is plotted as a line graph in Fig.~\ref{Fig3.fig}d. Although we are not in the strong SO coupling regime,
the distribution functions show a good agreement without fitting parameters. Based on the theory, $\langle
g^2\rangle=3/\lambda+\langle g_{o}^2\rangle$, where the first term is the spin contribution and $\langle
g_{o}^2\rangle$ is the orbital part~\cite{Matveev2000}. In the case of metallic nanoparticles the spin contribution
dominates the $g$-factor. In contrast, for InAs QDs the orbital part is expected to be the main contribution due to the
small effective mass and relatively large dot size. This is consistent with the large broadening of the distribution of
$g$. However RMT calculations~\cite{Matveev2000,Brouwer2000} focus mainly on the spin contribution. Thus a proper
estimation of the $g$-factor fluctuation for our InAs NW QD would require new model calculation, which takes into
account the orbital contribution induced by the QD geometry and the SO interaction of InAs.

In conclusion, we have analyzed the $g$-factor of discrete electron states of InAs nanowire based quantum dots. The
$|g|$ values were evaluated from the magnetic field splitting of zero bias Kondo ridges. The evaluated values show an
order of a magnitude large level-to-level fluctuation. Furthermore the $g$-factor can be tuned up to a factor of two
for the same charge state by gate electrodes. A possible explanation is drawn: the random nature of the wavefunction of
the discrete electron states generates different orbital contribution and spin-orbit correction to their $g$-factors.
The observed gate tunability of the $g$-factor of confined electrons can be used to drive the spin in and out of
resonance with a static magnetic field induced ESR condition, which could provide a fast and selective manipulation of
quantum dot based spin qubit.

We thank V.~Golovach, M.~Trif, J.~Paaske, D.~Zumb{\"u}l, A.~Eichler, M.~Weiss, G.~Zar{\'a}nd, L.~Borda, C.~M.~Marcus
and D.~Loss for discussions and C.~B.~S{\o}rensen for the help in the NW growth. This work has been supported by the
Swiss NSF, the NCCR on Nanoscale Science and the Danish Natural Science Research Council. S.~Csonka is a grantee of the
Marie Curie Fellowship.


\end{document}